\documentclass[11pt,showkeys,nofootinbib,notitlepage]{revtex4-1}

\usepackage{amsmath,amssymb,amsfonts,amsthm}
\usepackage{amsbsy} 
\usepackage{epsfig}
\usepackage{latexsym}
\input amssym.def
\input amssym.tex

\usepackage{hyperref}

\newcommand{\oarX}[1]{\href{http://arxiv.org/abs/#1}{{\ttfamily #1}}}
\newcommand{\arX}[1]{\href{http://arxiv.org/abs/#1}{{\ttfamily arXiv:#1}}}
\newcommand{\doin}[2]{\href{http://dx.doi.org/#1}{#2}}

\def\barr{\begin{array}}
\def\earr{\end{array}}

\def\ben{\begin{equation}}
\def\een{\end{equation}}
\def\bs{\begin{subequations}}
\def\es{\end{subequations}}
\def\bena{\begin{eqnarray}}
\def\eena{\end{eqnarray}}

\def\M{\mathcal{M}}
\def\be{\begin{equation}}
\def\ee{\end{equation}}
\def\bes{\begin{eqnarray}}
\def\ees{\end{eqnarray}}

\begin{document}

\title{Group field theory and its cosmology in a matter reference frame}

\author{Steffen Gielen}
\affiliation{School of Mathematical Sciences, University of Nottingham, University Park, Nottingham NG7 2RD, UK}
\email{steffen.gielen@nottingham.ac.uk}

\date{\today}


\begin{abstract}
While the equations of general relativity take the same form in any coordinate system, choosing a suitable set of coordinates is essential in any practical application. This poses a challenge in background-independent quantum gravity, where coordinates are not a priori available and need to be reconstructed from physical degrees of freedom. We review the general idea of coupling free scalar fields to gravity and using these scalars as a ``matter reference frame.'' The resulting coordinate system is harmonic, i.e.~it satisfies harmonic (de Donder) gauge. We then show how to introduce such matter reference frames in the group field theory approach to quantum gravity, where spacetime is emergent from a ``condensate'' of fundamental quantum degrees of freedom of geometry, and how to use matter coordinates to extract physics. We review recent results in homogeneous and inhomogeneous cosmology, and give a new application to the case of spherical symmetry. We find tentative evidence that spherically symmetric group field theory condensates defined in this setting can reproduce the near-horizon geometry of a Schwarzschild black hole.\end{abstract}


\maketitle

\section{Introduction}

Diffeomorphism symmetry is one of the cornerstones in the foundations of general relativity. A local choice of coordinate system is conventional and carries no physical meaning in itself; physical quantities (observables) must be invariant under diffeomorphisms, or changes of coordinate system. In physics language, diffeomorphisms form the gauge symmetries of general relativity. Fixing this vast gauge freedom is essential in practical applications of the theory, and key to extracting physical predictions. For spacetimes with isometries, using coordinates adapted to the symmetry simplifies the equations greatly; in cosmology, perturbations of a highly symmetric homogeneous and isotropic ``background'' spacetime must be classified as being gauge-invariant or gauge-dependent \cite{bardeen}.

As a gauge symmetry, diffeomorphism symmetry must also be taken into account when answering questions about the dynamical structure of general relativity, such as whether the initial value problem is well-posed. It turns out that a useful gauge choice for this is {\em harmonic gauge}, in which the Einstein equations have a well-posed initial value problem \cite{CYB}. Harmonic gauge (and its generalisations) can also be used to address the initial value problem in more general theories of gravity \cite{Reall}.

Harmonic gauge is defined by requiring the coordinates $x^\alpha$ to satisfy a wave equation,
\ben
\nabla^\mu\nabla_\mu x^\alpha\equiv\frac{1}{\sqrt{-g}}\partial_\mu\left(\sqrt{-g}g^{\mu\alpha}\right)=0\quad \Leftrightarrow\quad g^{\mu\nu}\Gamma^\alpha_{\mu\nu}=0\,.
\een
This statement is not generally covariant since the coordinates $x^\alpha$ do not form a vector, as it must be for this to be a gauge condition fixing diffeomorphism invariance. The use of harmonic coordinates has a long history in general relativity. Famously, their use was advocated by Fock as preferred coordinate systems in general relativity \cite{Fockbook}, in order for the structure of the theory to be as close as possible to that of special relativity (see e.g.~\cite{Fockdisc}). Proving the famous nonlinear stability result for Minkowski spacetime is easier in harmonic coordinates \cite{stability}. Harmonic gauge and generalised versions of it are also being successfully used in numerical relativity \cite{numGR}. While not all solutions of Einstein's equations admit a global harmonic coordinate system, such coordinates can clearly be used for a wide variety of situations of physical interest.

In quantum theory, one needs to verify that classical gauge symmetries are implemented without anomalies at the quantum level. In canonical loop quantum gravity (LQG), for instance, the question whether there is an anomaly-free representation of the Poisson algebra of spacetime diffeomorphisms, the hypersurface-deformation algebra \cite{HKT}, is the focus of ongoing work \cite{LQGanom}. In a diffeomorphism-invariant quantum theory it might then be possible to do calculations in a suitable gauge, as one can do for ${\rm U}(1)$ symmetry in quantum electrodynamics. However, making an appropriate gauge choice is only possible even in principle if the theory still contains enough diffeomorphism-variant structure, i.e.~gauge degrees of freedom. In canonical LQG, the starting point is the quantisation of continuum fields on a 3-dimensional manifold; but in the covariant (spin foam) formalism one abstracts from the underlying manifold and only works with combinatiorial structures \cite{CarloBook}. In this and other discrete settings for quantum gravity, the conventional notion of diffeomorphism symmetry is lost, and needs to be recovered in a continuum limit \cite{Bianca}. In general, (partial or full) background independence in such approaches implies that physical statements are formulated in relational terms, i.e.~concern physical degrees of freedom relative to another without reference to external structures \cite{CarloBook}. Background independence is conceptually important in implementing key principles of general relativity, but obstructs the possibility of extracting physical predictions which, as we mentioned, is usually done in suitable coordinates.

In a fully background-independent setting, where no background manifold structure supplying a conventional notion of coordinates is available, one needs to identify suitable {\em relational} ``coordinates'', i.e.~use the values of suitably chosen dynamical fields to parametrise the evolution of the remaining fields. If the theory under consideration only admits asking diffeomorphism-invariant questions, the conventional and convenient separation between coordinates and physical degrees of freedom breaks down. This paper illustrates various contexts in the group field theory approach in which scalar fields can serve as such relational coordinates. In general, the choice of particular matter fields as coordinates will only work locally, just like for any coordinate system, so that basic notions such as unitarity of the resulting quantum theory may not be easily definable. Furthermore, general covariance of the resulting quantum theory must still be demonstrated, independently of whether one chooses to use some matter degrees of freedom as coordinates.

To see all of this more concretely, we focus on the well-known example of a flat homogeneous, isotropic FLRW (Friedmann-Lema\^{i}tre-Robertson-Walker) universe with metric
\ben
{\rm d}s^2 = -N^2(t)\,{\rm d}t^2+a^2(t)\,{\rm d}x_3^2
\een
where ${\rm d}x_3^2$ is a fiducial flat metric on $\mathbb{R}^3$. Such a universe is foliated by homogeneous, isotropic hypersurfaces with flat intrinsic geometry; the geometry of these hypersurfaces relative to one another is parametrised by the scale factor $a(t)$. The dynamics of this universe are determined by the Friedmann equation
\ben
\left(\frac{1}{aN}\frac{{\rm d}a}{{\rm d}t}\right)^2 = \frac{8\pi G}{3}\rho
\label{friedm}
\een
where $\rho(t)$ is the energy density of matter and $N(t)$ is the lapse function. The lapse is an arbitrary function of coordinate time $t$, incorporating the freedom under time reparametrisations $t\mapsto t(t')$ under which $N(t)\mapsto N(t(t'))\left(\frac{{\rm d}t'}{{\rm d}t}\right)^{-1}$ so that (\ref{friedm}) is invariant. Proper time $N=1$ or conformal time $N=a$ are often useful coordinate choices for describing the evolution of the universe.

Standard Wheeler-DeWitt quantisation of the flat FLRW universe now leads to a wavefunction $\psi(a,\chi)$ where $\chi$ is a label for all matter degrees of freedom. There are operators corresponding to phase space functions, i.e.~functions of $a$ and $\chi$ and their momenta, but no operator corresponding to time $t$ which after all was just an arbitrary label. One can still ask relational questions, such as the evolution of matter fields $\chi(a)$ with the size of the universe,  but the convenience of working in proper time is no longer available. While proper time by definition always increases towards the future, this may not be true for physical degrees of freedom such as $a$ which may turn around (for $a$, this is a recollapse of the universe), leading to a breakdown of the parametrisation $\chi(a)$.

A popular way out of these technical complications is to introduce a matter field that can play the role of a ``matter clock.'' The standard candidate for this is a free and massless scalar field $\phi$ satisfying the Klein-Gordon equation
\ben
\nabla^\mu\nabla_\mu \phi = 0\,.
\een
In an FLRW universe, the Klein-Gordon equation reduces to
\ben
\frac{{\rm d}}{{\rm d}t}\left(\frac{a^3}{N}\frac{{\rm d}\phi}{{\rm d}t}\right) = 0\qquad \Rightarrow\;\frac{a^3}{N}\frac{{\rm d}\phi}{{\rm d}t}={\rm constant}\,.
\een
Assuming the constant is not zero, the evolution of $\phi$ is monotonic, and hence $\phi$ provides a good clock: it can never turn around. We can then write $\phi=\phi_0 T$, where the constant $\phi_0$ has dimensions of mass to make $T$ dimensionless, and use $T$ as ``scalar field time.'' The Klein-Gordon equation then becomes a harmonic coordinate condition on $T$, 
\ben
\nabla^\mu\nabla_\mu T = 0\qquad\Rightarrow\quad \frac{{\rm d}}{{\rm d}T}\left(\frac{a^3}{N}\right)=0\qquad \Rightarrow\quad N=\frac{a^3}{p}\,,\quad p\neq 0\,.
\label{harmoniccond}
\een
As expected, the harmonic condition fixes the lapse function; it is a gauge-fixing of time reparametrisations. For a flat FLRW universe only filled with this scalar field, the only remaining equation to solve is the Friedmann constraint equation
\ben
\left(\frac{{\rm d}a}{{\rm d}T}\right)^2 = \frac{4\pi G}{3}\phi_0^2\,a^2
\een
with two independent solutions
\ben
a(T)=a_0\exp\left(\pm \sqrt{\frac{4\pi G}{3}} \phi_0 T\right)\equiv a_0\exp\left(\pm \sqrt{\frac{4\pi G}{3}} \phi\right)
\label{solutions}
\een
corresponding to an expanding and contracting FLRW universe, respectively. The constant $p$ can be expressed in proper time $t_{{\rm p}}$ as
\ben
p=\frac{a^3}{N}=a^3\frac{{\rm d}T}{{\rm d}t_{{\rm p}}}\,,
\een
i.e.~$\phi_0 p$ corresponds to the usual conserved momentum of the scalar field.  The upshot of all this is that $\phi$ now serves as a preferred, harmonic ``matter time variable,'' with respect to which the equations can be solved. Predictions of a (so far unspecified) quantum theory for gravity coupled to a massless scalar can then be compared to the classical solutions (\ref{solutions}). This idea has been used since the early days of quantum cosmology \cite{QC}, and is fundamental in constructing the foundations of loop quantum cosmology \cite{LQC}, where quantisation ambiguities related to the choice of lapse function can be circumvented by requiring the time coordinate to be harmonic.

One of the main points we want to make is that this construction can be generalised. Namely, for a problem in quantum gravity that, after imposition of isometries, is effectively $k$-dimensional, we add $k$ free, massless scalars $\phi^i$ and introduce ``matter coordinates.'' They will satisfy harmonic conditions
\ben
\nabla^\mu\nabla_\mu x^i = 0\,,\quad \phi^i=\phi_0 x^i
\een
by virtue of the $\phi^i$ obeying the free Klein-Gordon equation. The $x^i$ are physical scalar fields which contribute to the energy-momentum tensor and thus backreact on spacetime. We saw this already in the FLRW case where the solution is not Minkowski spacetime, but an expanding or contracting universe. Nevertheless, their contribution can be negligible in certain regimes, as we will discuss below. We will show how the idea of adding massless scalar fields as matter coordinates is implemented in the setting of {\em group field theory}, where it has been employed to provide relational dynamics for an FLRW universe and, more recently, for perturbative inhomogeneities. Sections \ref{GFTref} to \ref{rods} are a review of these known results, as developed in \cite{GFcosmo,lowj,GFTinhomo}. In section \ref{GFTBH}, we then give a new application to the case of spherical symmetry. While ultimately unsatisfactory to describe the real universe (which does not contain free scalars, as far as we know) and limited in that the ``coordinates'' themselves are not properly quantised, we argue that this formalism can be useful in connecting quantum gravity to the vast experience with harmonic coordinates that we have in classical general relativity, and in overcoming some difficulties in extracting physical statements from a fundamentally background-independent framework. 

\section{Group field theory with reference matter}
\label{GFTref}

This section provides a short self-contained overview of the group field theory formalism for quantum gravity and matter, specifically its canonical formulation where one can define {\em group field theory condensates}, a proposal for states that describe a regime in which an effective macroscopic spacetime emerges dynamically. For general reviews of group field theory as a research programme for quantum gravity, see \cite{GFTrev}, for reviews of group field theory condensates and their application to (homogeneous) cosmology see \cite{GFTcondrev}.

In the canonical formulation of group field theory, the elementary degrees of freedom of geometry and matter are created as excitations of a bosonic, complex quantum field $\varphi$, itself defined on an abstract group manifold ({\em not} to be thought of as a spacetime). For models based on the Ashtekar-Barbero formalism for general relativity in terms of an ${\rm SU}(2)$ connection and with a scalar field coupled to gravity, $\varphi$ is a function
\ben
\varphi: {\rm SU}(2)^4 \times \mathbb{R} \rightarrow \mathbb{C}\,,\quad (g_I,\phi)\mapsto\varphi(g_I,\phi)\,.
\een
The arguments of $\varphi$ parametrise the elementary degrees of freedom of a ``chunk of space.'' In a formula, 
\begin{center}
\begin{picture}(320,100)
\put(40,50){$\hat\varphi^{\dagger}(g_1,g_2,g_3,g_4,\phi)\big|\emptyset\big\rangle=\Bigg|$}\put(290,50){$\Bigg\rangle$}
\put(222,52){$\bullet$}\put(220,100){\line(-1,-3){30}}\put(190,10){\line(3,1){90}}\put(280,40){\line(-1,1){60}}\put(190,10){\line(-1,1){30}}\put(160,40){\line(1,1){60}}
\put(165,60){$g_1$}\put(245,90){$g_2$}\put(245,40){$g_3$}\put(202,30){$g_4$}
\put(217,60){$\phi$}
\thicklines
\put(225,55){\line(-1,0){65}}\put(225,55){\line(1,1){45}}\put(225,55){\line(1,-1){50}}\put(225,55){\line(-1,-3){20}}
\end{picture}
\end{center}
defines a one-particle state created from the Fock vacuum $|\emptyset\rangle$; the state can be interpreted as an elementary tetrahedron, with four parallel transports $g_I$ of the gravitational ${\rm SU}(2)$ connection associated to links through its four faces, and with a label $\phi$ specifying the value of the (matter) scalar field. The role of $\hat\varphi^\dagger$ as an elementary creation operator derives from the fundamental commutation relations
\ben
\left[\hat\varphi(g_I,\phi),\hat\varphi^\dagger(g_I',\phi')\right]=\delta(\phi-\phi')\int_{{\rm SU}(2)}{\rm d}h\;\prod_{I=1}^4 \delta(g'_I h g_I^{-1})
\label{gftcommut}
\een
with all other commutators vanishing. The integration over ${\rm SU}(2)$ on the right-hand side ensures consistency with the ``gauge invariance'' property of the $\varphi$ field, namely its invariance under diagonal right multiplication
\ben
\varphi(g_1,\ldots,g_4,\phi)=\varphi(g_1h,\ldots,g_4h,\phi)\quad\forall h\in SU(2)
\label{gaugeinv}
\een
which is interpreted as an elementary ${\rm SU}(2)$ transformation acting on the central vertex.

One can generate states of arbitrary particle number $N$ by acting $N$ times with $\hat\varphi^\dagger(g_I,\phi)$ on $|\emptyset\rangle$ in the usual fashion; the GFT Hilbert space is a Fock space spanned by all of the subspaces corresponding to different values of $N$. Each such $N$-particle subspace can be associated to a ``graph'' of $N$ disconnected 4-valent open vertices. Graph connections can be introduced by integrating over the arguments corresponding to the links that one wants to glue: consider for instance a ``dipole state''
\ben
\int_{{\rm SU}(2)^4}{\rm d}^4h \;\hat\varphi^{\dagger}(h_1^{-1}g_1,h_2^{-1}g_2,h_3^{-1}g_3,h_4^{-1}g_4,\phi_1)\hat\varphi^{\dagger}(h_1,h_2,h_3,h_4,\phi_2)|\emptyset\rangle
\een
which can be interpreted as a state on two tetrahedra with all four faces identified, i.e.~a triangulation of the three-sphere. It is important to note that GFT states defined in this way do not ``know'' about the graph used to construct them: unlike in LQG, there is no unambiguous identification of an $N$-particle state with a graph or even a graph topology. For further discussion of the relation between the LQG and GFT Hilbert spaces see \cite{LQGGFT}, and \cite{Toy} for a discussion of the issue in a simplified toy model.

One now looks for states that can represent a macroscopic continuum geometry; it is clear that these cannot be states with a small number of excitations over the Fock vacuum, but that instead a large number of particles (often idealised to infinite) is needed.
 Being macroscopic means that expectation values of geometric observables such as the total 3-volume are required to be large compared to the fundamental (Planck) scale of the theory; semiclassical properties are imposed by focusing on field coherent states, similar to the coherent states used to describe Bose-Einstein condensates in condensed matter physics or macroscopic electromagnetic fields in quantum optics.

Such states are called {\em group field theory condensates}. The simplest example of a group field theory condensate is given by a mean-field coherent state of the form (up to normalisation)
\ben
|\sigma\rangle\propto\exp\left(\int_{{\rm SU}(2)^4\times \mathbb{R}}{\rm d}^4g\,{\rm d}\phi \;\sigma(g_1,g_2,g_3,g_4,\phi)\hat\varphi^{\dagger}(g_1,g_2,g_3,g_4,\phi)\right)|\emptyset\rangle
\label{coherent}
\een
where $\sigma$ is the analogue of the {\em condensate wavefunction} in condensed matter physics. The state $|\sigma\rangle$ is an eigenstate of the field operator $\hat\varphi$ and thus satisfies the symmetry breaking property
\ben
\langle \sigma|\hat\varphi(g_I,\phi)|\sigma\rangle = \sigma(g_I,\phi)
\een
which clearly distinguishes it from the Fock vacuum in which $\langle\emptyset| \hat\varphi(g_I,\phi) |\emptyset\rangle =0$. Moreover, in this state all normal-ordered $n$-point functions are simply products of $\sigma$ and its complex conjugate, e.g.
\ben
\langle \hat\varphi^{\dagger}(g_I,\phi)\hat\varphi^{\dagger}(g'_I,\phi')\hat\varphi(g''_I,\phi'')\hat\varphi(g'''_I,\phi''')\rangle = \overline{\sigma}(g_I,\phi)\overline{\sigma}(g'_I,\phi')\sigma(g''_I,\phi'')\sigma(g'''_I,\phi''')\,.
\label{npointf}
\een
The latter property implies an infinite tower of relations between all $n$-point functions, whose violation will signal the breakdown of this {\em mean-field approximation}. See \cite{GFCJHEP} for further background on the definition of these condensate states, their interpretation as macroscopic continuum geometries, and limits to the validity of the approximation. In group field cosmology, the validity of the mean-field approximation, at least away from the Planck regime and for large universes, is assumed as a working hypothesis. It has been shown, at least in simple GFT models, how a phase transition can lead to symmetry breaking and the formation of a condensate phase \cite{GFTphaset}; see also \cite{PithJo} for recent work on GFT phase transitions using Landau-Ginzburg theory.

The dynamics of such condensate states is now given by the analogue of what would be the Gross-Pitaevskii equation in condensed matter physics, i.e.~the expectation value of the quantum equation of motion. Using relations such as (\ref{npointf}), this expectation value reduces to the classical equation of motion for the GFT, to be satisfied by the mean field $\sigma$:
\ben
\langle\sigma|:\widehat{\frac{\delta S[\varphi,\overline{\varphi}]}{\delta \overline{\varphi}(g_I,\phi)}}:|\sigma\rangle = \frac{\delta S[\sigma,\overline{\sigma}]}{\delta \overline{\sigma}(g_I,\phi)} = 0\,.
\een
For a general GFT action
\ben
S[\varphi,\bar\varphi]=-\int_{{\rm SU}(2)^4\times \mathbb{R}}{\rm d}^4g\,{\rm d}\phi\,\bar\varphi(g_I,\phi)\mathcal{K}\varphi(g_I,\phi)+\mathcal{V}[\varphi,\bar\varphi]
\label{akshn}
\een
given in terms of a quadratic term including a kinetic operator $\mathcal{K}$ and a general potential $\mathcal{V}$, the mean field equation of motion becomes
\ben
\mathcal{K}\sigma(g_I,\phi)-\frac{\delta \mathcal{V}[\sigma,\overline{\sigma}]}{\delta \overline{\sigma}(g_I,\phi)} = 0\,.
\label{GPeq}
\een
There are various directions towards choosing suitable forms for $\mathcal{K}$ and $\mathcal{V}$. If GFT is used to define dynamics for loop quantum gravity, they are chosen so that the Feynman amplitudes of the GFT correspond to the amplitudes of a given spin foam model for the boundary data given by GFT Fock states as defined above. Such a choice is possible in quite some generality, so that there is a one-to-one correspondence of spin foam models and GFT actions \cite{RovelliReis}. In this construction, the kinetic term is typically trivial, i.e.~$\mathcal{K}$ is simply a constant in (\ref{akshn}). Studying renormalisation in GFT requires the introduction of a notion of scale which is done by adding a group Laplacian into the kinetic term; such a nontrivial kinetic term would be generated through radiative corrections \cite{GFTrenorm}. Such results then suggest that more general forms of $\mathcal{K}$ which do include derivatives with respect to all arguments of $\varphi$ should be considered. In any case, the potential $\mathcal{V}$ typically has a combinatorial type of nonlocality corresponding to simplicial gluing of the fundamental GFT building blocks into higher-dimensional structures. In this paper we present results applicable to a wide range of models under the given assumptions, and no specific forms of $\mathcal{K}$ or $\mathcal{V}$ will be used.

In general, since $\mathcal{K}$ and $\mathcal{V}$ can take rather complicated forms, the task of finding explicit solutions to (\ref{GPeq}) is difficult. To get a first understanding of the effective dynamics of GFT condensates in spacetime terms, one now makes two types of approximations. The first, as developed in \cite{GFcosmo}, is to expand the kinetic operator $\mathcal{K}$ in derivatives with respect to $\phi$,
\ben
\mathcal{K} = \mathcal{K}_0 + \mathcal{K}_2 \frac{\partial^2}{\partial\phi^2} + \ldots
\label{expansion}
\een
and truncate after the second derivative with respect to $\phi$. Notice that the ``coefficients'' in this expansion $\mathcal{K}_0$ and $\mathcal{K}_2$ are still differential operators with respect to the ${\rm SU}(2)$ variables $g_I$. For GFT models for a massless, free scalar ``matter clock,'' the kinetic operator $\mathcal{K}$ has no explicit dependence on $\phi$ and the expansion (\ref{expansion}) only contains even derivatives. This is because the GFT action is invariant under shifts $\phi\mapsto\phi+\phi_0$ and time reversal transformations $\phi\mapsto -\phi$, the symmetries of such a clock field.

A second approximation often made is to consider a weak-coupling limit in which the potential $\mathcal{V}$ is neglected. To some extent this approximation reflects a need for self-consistency: the mean-field coherent state $|\sigma\rangle$, in which all GFT quanta are uncorrelated, must to be a good approximation to an exact physical state of the theory. Indeed, for Bose-Einstein condensates such states describe a weakly coupled regime and are not suitable at strong coupling. Since the potential $\mathcal{V}$ includes higher powers of $\sigma$ relative to the kinetic term, it becomes more and more relevant with a growing particle number (which grows as $|\sigma|^2$), so that the free approximation can be expected to be valid in a mesoscopic regime in which the particle number is large but not too large. Interaction terms can be included into the cosmological analysis, as in \cite{KCLgroup} where they become important at late times and can lead to a recollapse of the universe analogous to a negative cosmological constant.

The weak-coupling limit is closely related to the choice of coherent state $|\sigma\rangle$, and its failure is often seen in a breakdown of this mean-field treatment. Strong interactions typically correspond a dynamical regime in which higher $n$-point functions no longer approximately follow (\ref{npointf}).

\section{Effective cosmological dynamics}

Within this mean-field approximation for GFT condensates, one now computes expectation values of geometric observables in order to derive an effective cosmological dynamics from the GFT Gross-Pitaevskii equation (\ref{GPeq}). The simplest such observable is the total 3-volume of the universe at a given value $\phi$ of the  matter clock field, which corresponds to the GFT Fock space operator \cite{GFcosmo}
\ben
\hat{V}(\phi) = \int_{{\rm SU}(2)^4\times {\rm SU}(2)^4}{\rm d}^4g\,{\rm d}^4h\;V(g_I,h_I)\hat\varphi^{\dagger}(g_I,\phi)\hat\varphi(h_I,\phi)
\label{volumeop}
\een
where $V(g_I,h_I)$ are the matrix elements of the LQG volume operator, calculated between two spin network states $|g_I\rangle$ and $|h_I\rangle$ on a single open 4-valent vertex. The expectation value $V(\phi)\equiv\langle \hat{V}(\phi)\rangle$ then satisfies an effective Friedmann equation, which can be compared to the equations for $a(T)$ (parametrised by matter time $T$) which appeared in the classical discussion around (\ref{solutions}).

It turns out that the simple approximations used up to this point already allow the extraction of physically reasonable effective dynamics from GFT condensates \cite{GFcosmo}. In the mean-field approximation and weak-coupling limit, the GFT Gross-Pitaevskii equation (\ref{GPeq}) becomes simply
\ben
\left(\mathcal{K}_0 + \mathcal{K}_2 \frac{\partial^2}{\partial\phi^2}\right)\sigma(g_I,\phi)=0\,.
\label{simpleGP}
\een
We should stress again that the weak-coupling limit, which is here made for simplicity and self-consistency, is an assumption of what follows. We will see that interesting phyiscally relevant cosmological dynamics can already be derived within this simplest possible approximation.
 
As is often the case in LQG, the equation (\ref{simpleGP}) is best analysed in a Peter-Weyl expansion of functions on ${\rm SU}(2)$ into irreducible representations. One can restrict this expansion to {\em isotropic} modes, the modes characterising building blocks that are themselves isotropic, with geometric interpretation as equilateral tetrahedra. The Peter-Weyl expansion then takes the form
\ben
\sigma(g_I,\phi)=\sum_{j\in \frac{\mathbb{N}_0}{2}} \sigma_j(\phi)\,{\bf D}^j(g_I)
\een
where only a single spin $j$ labelling an irreducible representation of ${\rm SU}(2)$ appears. Notice that all $g_I$ dependence is now in  the ${\bf D}^j(g_I)$ which are an appropriate convolution of four Wigner $D$-matrices with intertwiners, taking care of ${\rm SU}(2)$ gauge invariance. The restriction to a single $j$ in the expansion is purely for convenience: isotropic modes are sufficient to capture the dynamics of an FLRW universe. Further modes with different $j$ labels for the four faces of a tetrahedron can be included and their effects on the effective dynamics can be studied \cite{anisopaper}.

For usual GFT actions, $\mathcal{K}$ can contain derivatives but no explicit $g$-dependence. The Peter-Weyl decomposition then leads to a decoupling of (\ref{simpleGP}) into separate equations for each $j$, of the form
\ben
\left(-B_j+A_j \frac{\partial^2}{\partial\phi^2}\right)\sigma_j(\phi)=0
\label{jGP}
\een
where $A_j$ and $B_j$ are now $j$-dependent couplings which depend on the kinetic term of the original GFT action; in particular, in the commonly assumed situation where $\mathcal{K}_0$ and $\mathcal{K}_2$ are general functions of Laplace-Beltrami operators with respect to the $g_I$, we could define
\ben
\mathcal{K}_0{\bf D}^j(g_I) =: -B_j{\bf D}^j(g_I)\,,\quad \mathcal{K}_2{\bf D}^j(g_I) =: A_j{\bf D}^j(g_I)
\label{coefficients}
\een
where each Laplacian (acting on one of the $g_I$) would contribute an eigenvalue $-j(j+1)$. It is immediate to see that the solutions to (\ref{jGP}) have the form
\ben
\sigma_j(\phi)=\alpha_j^+\exp\left(\sqrt{\frac{B_j}{A_j}}\phi\right) + \alpha_j^-\exp\left(-\sqrt{\frac{B_j}{A_j}}\phi\right)\,.
\label{salushn}
\een
For the simplest case in which only a single mode $j=j_0$ is excited, the total 3-volume then asymptotes to
\ben
V(\phi)=\sum_j V_j |\sigma_j(\phi)|^2 = V_{j_0}|\sigma_{j_0}(\phi)|^2 \stackrel{\phi\rightarrow\pm\infty}{\sim} V_{j_0} |\alpha_j^\pm|^2 \exp\left(2\sqrt{\frac{B_{j_0}}{A_{j_0}}}\phi\right)
\een
at late and early times (as given by $\phi$), where we assume that $\frac{B_j}{A_j}>0$. This is a bounce solution; it interpolates between the classical expanding and contracting solutions (\ref{solutions}) of the classical Friedmann dynamics if we identify $B_{j_0}/A_{j_0}=:3\pi G$ as the ``emergent'' Newton's constant. One can show that a singularity ($V(\phi)=0$ for some $\phi$) is only possible for special initial conditions, that for a more general configuration in which many $j$ are excited, a single $j$ will dominate asymptotically in a wide range of GFT models, and that the single $j=j_0$ case reproduces bounce dynamics very similar to the ``improved dynamics'' of loop quantum cosmology \cite{GFcosmo,lowj}. Together, these results show the potential of GFT condensates to provide a realistic dynamics for cosmology, at least in the very simplest case of a flat FLRW universe filled only with a massless scalar clock field. Most of these results can already be found in the simpler toy model of \cite{Toy}.

\section{Beyond homogeneity: adding rods}
\label{rods}

It is desirable to extend this framework beyond the spatially homogeneous case, in order to make contact with more realistic descriptions of cosmology in which inhomogeneities are present, and to explore the possibility of GFT condensates describing other geometries of physical interest, in particular black holes. If isometries are present, the problem is still effectively lower-dimensional, but it should also be possible to include an arbitrary dependence on spacetime coordinates.

As we discussed in the introduction, a possible way forward is now to include not one but $k$ scalar fields $\phi^i$ that can serve as material clocks and rods. Taking these to be massless, free scalars, they will each satisfy the Klein-Gordon equation, here interpreted as a harmonic coordinate condition
\ben
\nabla^\mu\nabla_\mu \phi^i = 0\,.
\een
Including them into the GFT formalism is entirely straightforward: the GFT field is now a function
\ben
\varphi: {\rm SU}(2)^4 \times \mathbb{R}^k \rightarrow \mathbb{C}\,,\quad (g_I,\phi^J)\mapsto\varphi(g_I,\phi^J)
\een
on the extended configuration space ${\rm SU}(2)^4 \times \mathbb{R}^k$ of the tetrahedron. Instead of a single matter field label $\phi$, there are now $k$ labels $\phi^J$ attached to the vertex of each tetrahedron, corresponding to the readings of all clock and rod fields. Dynamics are still formulated in relational terms, now using all $k$ fields to distinguish spacetime points in the emergent spacetime description (i.e.~in the condensate phase). This approach to extending the GFT formalism in order to describe inhomogeneous geometries was put forward in \cite{GFTinhomo} in order to include perturbative inhomogeneities; see also \cite{GFTsepuni} for a related cosmological application.\footnote{For an orthogonal approach to describing spherically symmetric spacetimes with GFT condensates see \cite{GFTgencond}.} The use of material reference frames in quantum gravity is of course not new. Many authors have used dust matter to this effect, starting from Brown and Kucha\v{r} \cite{brownkuch}; generalisations of the Brown-Kucha\v{r} model have been applied to LQG \cite{LQGdust}. The general idea of using convenient matter scalar variables as coordinates dates at least back to DeWitt \cite{DeWitt}; see also \cite{GHM} for an extended discussion. There are also studies in LQG based on massless scalar fields \cite{LQGscalars}. As we continue to argue in this section, adopting material reference systems in GFT leads to new ways of extracting effective spacetime dynamics from a fundamental, non-perturbative approach to quantum gravity.

In this extended setting with $k$ scalars, the different approximations, in particular the use of mean-field coherent states to describe weakly interacting GFT condensates, go through in complete analogy to the case of a single clock field. Again one ends up with a Gross-Pitaevskii-type equation
 \ben
\mathcal{K}\sigma(g_I,\phi^J)-\frac{\delta \mathcal{V}[\sigma,\overline{\sigma}]}{\delta \overline{\sigma}(g_I,\phi^J)} = 0\,.
\label{GPeq2}
\een
For concreteness, let us now assume $k=4$, i.e.~one has coupled sufficiently many fields to have a linearly independent set of clocks and rods available at each point in spacetime.\footnote{Technically, the condition for this would be $\det(\partial_\mu\phi^J)\neq 0$ in an arbitrary coordinate system, which is generically satisfied. In the homogeneous cosmological case this non-degeneracy condition is $p\neq 0$, see (\ref{harmoniccond}).} The expansion of $\mathcal{K}$ in derivatives now takes the form
\ben
\mathcal{K} = \mathcal{K}_0 + \mathcal{K}_2 \frac{\partial^2}{\partial(\phi^0)^2}  + \tilde{\mathcal{K}}_2 \sum_{i=1}^3\frac{\partial^2}{\partial(\phi^i)^2} + \ldots
\label{expansion2}
\een
assuming the following symmetries of the GFT action: invariance under shifts in any of the $\phi^J$; parity and time reversal $\phi^J\mapsto -\phi^J$; and arbitrary rotations of the rods $\phi^i\mapsto {O^i}_j\phi^j$. These are symmetries of the classical action for massless scalar fields, and they are reasonable conditions for any material coordinate system assuming that translation invariance, parity, time reversal and isotropy of space are indeed fundamental symmetries of Nature (as we assume they are). We do not assume Lorentz invariance to be fundamental but one could do so.

Notice that this extended framework with four massless scalars coupled to gravity includes all of the homogeneous GFT condensates discussed previously as solutions: these are given by homogeneous condensate wavefunctions of the form
\ben
\sigma(g_I,\phi^J)\equiv \sigma^0(g_I,\phi^0)\,.
\label{homo}
\een
One can define both homogeneous and more general inhomogeneous condensates as solutions for the same GFT action.

If we again use a Peter-Weyl decomposition and again, for simplicity, restrict the mean field to only include isotropic modes, the generalisation of the effective dynamics (\ref{jGP}) is now
\ben
\left(-B_j+A_j \frac{\partial^2}{\partial(\phi^0)^2}+C_j\sum_{i=1}^3\frac{\partial^2}{\partial(\phi^i)^2}\right)\sigma_j(\phi^J)=0\,,
\label{newGP}
\een
with the coefficients $B_j$, $A_j$ and $C_j$ arising from a mode decomposition of the kinetic operator $\mathcal{K}$ in complete analogy to (\ref{coefficients}). Again it is obvious that if the mean field does not depend on the ``rod'' fields $\phi^i$, the equation reduces to the previous homogeneous case. One can perturb around exact homogeneity, 
\ben
\sigma_j(\phi^J) = \sigma_j^0(\phi^0)\left(1+\epsilon \,\psi_j(\phi^J)\right)\,;
\een
the field $\psi_j(\phi^J)$ then describes condensate perturbations analogous to phonons in a real Bose-Einstein condensate. Using phonons to perturb GFT condensates around exact homogeneity had been proposed already in \cite{GFCJHEP}, but their effective spacetime interpretation was not apparent without matter fields that could be used as relational rods. These phonons are then perturbations in the mean field, i.e.~at the level of expectation values, interpreted directly as deviations from homogeneity in the effective spacetime geometry.

The proposal of \cite{GFTinhomo} is to go one step further and consider quantum fluctuations in the local volume as the seeds of cosmological inhomogeneities. This is analogous to what happens in inflation, where one computes the power spectrum of quantum fluctuations in an exactly homogeneous quantum state (rather than a quantum state on a classically perturbed geometry). Such fluctuations must always be there due to quantum uncertainty; they then ``freeze out'' and are converted into the pattern of classical inhomogeneities observed in the CMB \cite{mukhanov}. Similarly, quantum uncertainty sets a lower bound on deviations from exact homogeneity in a GFT condensate \cite{oldinhomo}.

In the extended GFT formalism with clock and rod fields, a local volume element operator can be defined by
\ben
\hat{V}(\phi^J)=\int_{{\rm SU}(2)^4\times {\rm SU}(2)^4}{\rm d}^4g\,{\rm d}^4h\;V(g_I,h_I)\hat\varphi^\dagger(g_I,\phi^J)\hat\varphi(h_I,\phi^J)\,.
\label{volumeelement}
\een
Notice the similarity with (\ref{volumeop}); $\hat{V}(\phi^J)$ now corresponds to the volume element at the ``spacetime point'' specified by the values $\phi^J$ of the clock and rod fields. (To be precise, $\hat{V}(\phi^J)$ is a density and a finite volume is obtained from $\hat{V}(\phi^J)\delta^4\phi^J$ where $\delta^4\phi^J$ is an infinitesimal volume in field space.) This is precisely the quantity needed to characterise scalar perturbations in cosmology, which are perturbations in the local volume element of the spacetime geometry. Physically, since in this formalism scalar fields need to be coupled to gravity to serve as material clocks and rods, one would expect scalar perturbations to be present. 

As shown in \cite{GFTinhomo}, for an exactly homogeneous mean field (\ref{homo}) one finds
\ben
\langle \hat{V}(\phi^0,k_i) \hat{V}(\phi'^0,k'_i)\rangle - \langle \hat{V}(\phi^0,k_i)\rangle\langle \hat{V}(\phi'^0,k'_i)\rangle = (2\pi)^3\delta^3(k_i+k'_i)\delta(\phi^0-\phi'^0) \sum_j V_j^2\,|\sigma_j^0(\phi^0)|^2 \label{powersp}
\een
where a Fourier transform has been performed from the rod fields $\phi^i$ to their wavenumbers $k^i$ in order to bring the power spectrum to the usual Fourier space form. The right-hand side is scale invariant with respect to the rod wavenumbers $k^i$ and the relative amplitude of these inhomogeneities is naturally small: it scales as $1/N$, where $N$ is the number of quanta in the condensate, which is the typical relative size of fluctuations in a condensate. These results are the first step towards developing a new formalism for generating cosmological perturbations purely from quantum fluctuations in a single quantum theory for gravity and matter. 

When comparing this formalism with usual cosmology one must take into account that the power spectrum is given with respect to wavenumbers $k^i$ for the harmonic coordinate system given by the scalar fields; the study of how the initial quantum fluctuations given in (\ref{powersp}) evolve dynamically and are converted into classical perturbations is then the next step. In usual cosmological perturbation theory, harmonic gauge is of advantage when studying bouncing cosmologies, where the propagation of perturbations through a bounce can be well defined in this gauge \cite{harmonicBounce}. Since GFT condensates provide an example of a bouncing cosmology, such coordinates are particularly useful.

\section{Clocks and rods for a black hole}
\label{GFTBH}

Apart from adding inhomogeneities to an FLRW universe to do cosmology, a second application that goes beyond spatial homogeneity is using GFT condensates with material clock and rod fields in order to describe black holes. The results in this section are new, and may pave the way towards more detailed investigations, as we shall discuss at the end.

Let us first summarise the corresponding classical theory. Here, one has the Einstein equations
\ben
R_{\mu\nu}-\frac{1}{2} R g_{\mu\nu}=8\pi G \sum_{J} T^J_{\mu\nu}
\een
where $T^J_{\mu\nu}$ is the energy-momentum tensor for the scalar field $\phi^J$ and these are all massless, free, non-interacting scalars. The $\phi^J$ themselves, as we already mentioned several times, satisfy the free Klein-Gordon equation, and the point is that the values of the $\phi^J$ are themselves used as coordinates: we make the (locally possible) gauge choice $\phi^J=\phi_0 x^J$ where $\phi_0$ has units of mass to make $x^J$ dimensionless. In this matter reference frame the energy-momentum tensors take the particularly simple form (no sum over $J$ here)
\ben
T^J_{\mu\nu}=\partial_\mu\phi^J\partial_\nu\phi^J-\frac{1}{2} g_{\mu\nu}(\partial\phi^J)^2=\phi_0^2\left({\delta_\mu}^J {\delta_\nu}^J - \frac{1}{2} g_{\mu\nu}g^{JJ}\right)\,.
\een
The right-hand side of Einstein's equations is thus multiplied by the dimensionless factor $8\pi G \phi_0^2\sim(\phi_0/M_{{\rm P}})^2$ which one may attempt to treat as a ``small number'' in a perturbative expansion; notice however that the value of $\phi_0$ is arbitrary since one can always perform a global rescaling without changing the harmonic coordinate condition.

We are now interested in spherically symmetric condensate solutions in GFT. These will in general depend on two directions, a timelike and a radial coordinate, corresponding to a dependence on two reference fields $\phi^T$ and $\phi^R$. Spherical symmetry means that we can assume that even if two further rod fields are present, the GFT mean field is independent of these. We do not change the fundamental definition of a condensate through mean-field coherent states such as (\ref{coherent}), which we consider as candidate GFT configurations for general macroscopic, effective continuum configurations. In particular, nothing in the definition makes reference to concepts such as an event horizon, which one would hope emerge from the effective semiclassical GFT dynamics.

At the classical level, such condensates should correspond to spherically symmetric solutions of general relativity coupled to two free scalar fields and written in harmonic coordinates 
\ben
\phi^T \equiv \phi_0 T\,, \quad \phi^R \equiv \phi_0 R\,.
\een
Without further assumptions, finding such solutions analytically looks like a rather daunting task. 

A simpler and perhaps more natural first question would be whether there is an analogue of the Schwarzschild solution in GFT. The Schwarzschild metric is a vacuum solution of general relativity, whereas in our formalism we require scalar fields with nontrivial stress-energy (i.e.~not simply constant fields) in order for them to be viable components of a matter reference frame; but this stress-energy can be arbitrarily small as long as it is non-zero. This suggests the existence of relevant black hole solutions which are in the vicinity of the black hole given by small perturbations of the Schwarzschild metric, although they will globally differ (in particular, they will not be asymptotically flat but asymptote to an expanding or contracting FLRW metric).

As a starting point, it may then be useful to work in vacuum, where we know that any spherically symmetric solution to the Einstein equations is static. Of course, the solution is the Schwarzschild metric, but it is instructive to derive it in terms of a harmonic radial coordinate. Start with a general spherically symmetric and static metric ansatz of the form
\ben
{\rm d}s^2 = -N^2(R)\,{\rm d}T^2+A^2(R)\,{\rm d}R^2 + B^2(R)\,{\rm d}\Omega^2
\een
where ${\rm d}\Omega^2$ is the round metric on the two-sphere, corresponding to the orbits of the isometry group ${\rm SO}(3)$. The harmonic gauge condition then becomes
\ben
\frac{{\rm d}}{{\rm d}R}\left(\frac{NB^2}{A}\right)=0\quad \Rightarrow\quad N=\frac{kA}{B^2}\,,\quad k\neq 0\,,
\een
again fixing the lapse function.

Because we assume the metric to be static, the problem is effectively still one-dimensional, allowing us to solve the equations analytically. The two nontrivial Einstein equations are
\bena
A(R)^3 + 2A'(R)B(R)B'(R)-A(R)\left(B'(R)^2+2B(R)B''(R)\right)&=&0\,,\nonumber
\\A(R)^2+B'(R)^2-B(R)B''(R)&=&0\,.
\eena
The first equation is a first-order ODE for $A(R)$ with general solution
\ben
A(R)=\pm\sqrt{\frac{B(R)}{B(R)+C}}\,B'(R)\,;
\een
the second equation then becomes
\ben
(C+B(R))B''(R)-\left(2+\frac{C}{B(R)}\right)B'(R)^2=0
\een
with general solution
\ben
B(R)=\frac{C}{e^{-\lambda (R+R_0)}-1}\,.
\label{bsol}
\een
The constants $R_0$ and $\lambda$ can be chosen freely ($\lambda$ could always be absorbed into a redefinition of $\phi_0$), and we set $R_0=0$ and $\lambda=1$. The metric then takes the form
\ben
{\rm d}s^2 = -\frac{k^2}{C^2}e^{-R}\,{\rm d}T^2+\frac{C^2\,e^{-R}}{\left(e^{-R}-1\right)^4}\,{\rm d}R^2 + \frac{C^2}{\left(e^{-R}-1\right)^2}\,{\rm d}\Omega^2\,.
\label{metric}
\een
In order to recognise that this is indeed the Schwarzschild metric, introduce a new (non-harmonic!) radial coordinate
\ben
r=\frac{C}{1-e^{-R}}\,,\quad e^{-R}=1-\frac{C}{r}
\een
so that spheres of constant $(T,r)$ have area $4\pi r^2$. This brings the metric into the more familiar form
\ben
{\rm d}s^2 = -\frac{k^2}{C^2}\left(1-\frac{C}{r}\right){\rm d}T^2+\left(1-\frac{C}{r}\right)^{-1}{\rm d}r^2 + r^2\,{\rm d}\Omega^2\,.
\een
We now recognise the constant $C$ as the Schwarzschild radius $r_{{\rm S}}=2MG$ of the black hole. For $C\ge 0$ the radial coordinate $r$ should also be positive, which implies that $R>0$ in our original coordinate system; positive values of $R$ cover the exterior Schwarzschild solution, where $R=0$ is asymptotic infinity and $R=\infty$ corresponds to the horizon. Negative values of  $R$ (and $r$) would correspond to a Schwarzschild solution with negative mass, which describes a naked singularity; indeed the harmonic $R$ coordinate can then be extended all the way to the singularity at $R=-\infty$.

In order to compare this classical black hole solution to condensate solutions in GFT, we need to identify a suitable geometric observable in the Schwarzschild spacetime and express it in the relational coordinates $(T,R)$. As in the case of cosmology, a good candidate for this is the 3-volume of a region in spacetime, defined by a coordinate patch in the matter coordinates. Consider a shell between radial coordinate values $R$ and $R+\delta R$; its 3-volume is
\ben
V_{{\rm shell}}(R)=\int {\rm d}^3 x\;\sqrt{h} = 4\pi r_{{\rm S}}^3\int\limits_R^{R+\delta R} {\rm d}s\;\frac{e^{-s/2}}{\left(e^{-s}-1\right)^4}\simeq 4\pi r_{{\rm S}}^3 \,\delta R\frac{e^{-R/2}}{\left(e^{-R}-1\right)^4}\,.
\een
In the vicinity of the black hole horizon where $e^{-R}\ll 1$ this is well approximated by
\ben
V_{{\rm shell}}(R)\simeq 4\pi r_{{\rm S}}^3 \,e^{-R/2}\,\delta R\,.
\label{nearhor}
\een
The $R$ coordinate system breaks down at $R=\infty$, just as the Schwarzschild coordinate system breaks down at $r=2MG$. For the negative mass Schwarzschild solution with $e^{-R} \gg 1$, we would similarly approximate
\ben
V_{{\rm shell}}(R)\simeq 4\pi r_{{\rm S}}^3 \,e^{7R/2}\,\delta R\,.
\een

We can now compare this behaviour to that of spherically symmetric GFT condensates. In the approximations considered in this paper, such condensates satisfy (\ref{newGP}) where we assume the GFT mean field to only depend on one clock and one rod field, i.e.
\ben
\left(-B_j+A_j\frac{\partial^2}{\partial (\phi^T)^2}+C_j\frac{\partial^2}{\partial (\phi^R)^2}\right) \sigma_j(\phi^T,\phi^R)=0\,.
\een
For a static solution, we should also assume the field to be independent of $\phi^T$. Then
\ben
\left(-B_j+C_j\frac{\partial^2}{\partial (\phi^R)^2}\right) \sigma_j(\phi^R)=0\quad\Rightarrow\quad \sigma_j(\phi^R)=a_j\exp\left(\sqrt{\frac{B_j}{C_j}}\phi^R\right)+b_j\exp\left(-\sqrt{\frac{B_j}{C_j}}\phi^R\right)\,.
\een
The equation of motion and its solutions take the same functional form as for spatially homogeneous condensates (\ref{jGP})-(\ref{salushn}), because we are again dealing with an effectively one-dimensional problem. However, a different coupling $C_j$ appears, which can be different from the $A_j$ relevant for cosmology.
\\For a single spin $j=j_0$ and only considering solutions for which $a_{j_0}=0$, we find that the GFT shell volume operator $\hat{V}_{{\rm shell}}(\phi^R) := \hat{V}(\phi^R)\delta\phi^R$ has an expectation value
\ben
V_{{\rm shell}}(\phi^R) =\langle \hat{V}_{{\rm shell}}(\phi^R)\rangle = V_{j_0} |b|^2\exp\left(-2\sqrt{\frac{B_{j_0}}{C_{j_0}}}\phi^R\right)\delta \phi^R
\label{Ashellvolume}
\een
which, upon setting $\sqrt{\frac{B_{j_0}}{C_{j_0}}}\phi_0 = \frac{1}{4}$, reproduces the classical black hole near-horizon behaviour (\ref{nearhor}). This condition leaves the values of $B_j$ or $C_j$ unconstrained, since the value of $\phi_0$ is arbitrary as we noted below (\ref{bsol}).\footnote{It might appear puzzling that the correct near-horizon behaviour does not depend on the fundamental GFT couplings, but this is at is should be: we are trying to match with a vacuum solution of general relativity, for which the value of Newton's constant likewise plays no role.} Similarly, solutions for a single spin in which $b_{j_0}=0$ give
\ben
V_{{\rm shell}}(\phi^R) =V_{j_0} |a|^2\exp\left(2\sqrt{\frac{B_{j_0}}{C_{j_0}}}\phi^R\right)\delta \phi^R
\label{GFTshellvolume}
\een
and we can now set $\sqrt{\frac{B_{j_0}}{C_{j_0}}}\phi_0 = \frac{7}{4}$ to find the near-singularity behaviour of a negative mass Schwarzschild solution (recall that the choice of $\phi_0$ is arbitrary, so that fulfilling such a condition is always possible). There is no evidence for singularity resolution in this case, as (\ref{GFTshellvolume}) never deviates from the classical expression for the shell volume. Given that the negative mass Schwarzschild solution has a naked singularity, one may not want to see it resolved in quantum gravity; we should however point out that the solutions we consider display extreme fine-tuning, since one has to set either $a_{j_0}=0$ or $b_{j_0}=0$. Similarly finely tuned solutions also exist in the cosmological context (\ref{salushn}), and they are exactly the solutions in which there is {\em no} bounce.

The coefficients $|a|^2$ or $|b|^2$ correspond to the number of GFT quanta in these states, which must then be given by $4\pi r_{{\rm S}}^3/V_{{j_0}}$ to describe a black hole with Schwarzschild radius $r_{{\rm S}}$.

For the physically relevant case of a positive mass Schwarzschild solution, we only have the exterior solution at our disposal, since the harmonic radial coordinate $R$ only reaches the horizon. This is a drawback of having to restrict to harmonic coordinates in our construction. It then does not seem straightforwardly possible to make statements about singularity resolution in GFT black holes; rather, a comparison to recent work on effective black-hole metrics in LQG (e.g.~\cite{LQGbh}) can only be done for the exterior. Here it is noteworthy that the exterior Schwarzschild solution can be exactly identical to that of general relativity even if LQG-type corrections are included \cite{effective}. A second question whose answer would require access to the interior is whether one can calculate entanglement entropy in a GFT condensate state across the putative black hole horizon in order to recover an area law, as was successfully done in \cite{GFTgencond}. If there was a GFT coherent state of the form (\ref{coherent}) describing both the exterior and interior of a black hole, the associated entanglement entropy would presumably be equal to that of the GFT Fock vacuum, since this is the case for coherent states in a general scalar field theory \cite{ententropy}. Effective studies of quasi-normal modes or perhaps even black hole evaporation might be possible using only the exterior solution.

The presented evidence for suggesting that spherically symmetric GFT condensates can match the near-horizon behaviour of a black hole in general relativity is tentative as it only uses the shell volume as a single observable and not e.g.~the area of a constant $(T,R)$ surface of codimension two or the extrinsic or intrinsic curvature of a hypersurface in the effective geometry, which would give further insights into the spacetime interpretation of this condensate mean field. While encouraging in reproducing the correct classical limit, the results are also less convincing than in the FLRW case as they rely on fine-tuning in the choice of condensate solutions; such fine-tuning should be justified by other considerations independent from enforcing a match with classical general relativity. Nevertheless, this is the first foray into quantitative studies of effective black hole geometries in GFT, and should be seen as the starting point for further research aimed at extending the results found for GFT cosmology to GFT black holes, and then other situations of physical interest. We have already seen how harmonic matter coordinates can prove to be a valuable tool for investigating general inhomogeneous geometries in full quantum gravity, outside of the usual symmetry-reduced models in which symmetry reduction is applied before quantisation.

{\em Acknowledgements.} --- I would like to thank Suddhasattwa Brahma, Roberto Percacci and two referees for helpful comments. This research was funded by the Royal Society under a Royal Society University Research Fellowship (UF160622) and a Research Grant for Research Fellows (RGF\textbackslash R1\textbackslash 180030).

\end{document}